\pdfoutput=1

\documentclass[twoside,twocolumn]{article}

\usepackage{blindtext} % Package to generate dummy text throughout this template 
\usepackage{lmodern} % Use the Palatino font
\usepackage[T1]{fontenc} % Use 8-bit encoding that has 256 glyphs
\usepackage{microtype} % Slightly tweak font spacing for aesthetics
\usepackage{chemfig}

\usepackage[english]{babel}

\usepackage[hmarginratio=1:1,top=32mm,columnsep=20pt]{geometry} % Document margins
\usepackage[hang, small,labelfont=bf,up,textfont=it,up]{caption} % Custom captions under/above floats in tables or figures
\usepackage{booktabs} % Horizontal rules in tables

\usepackage{lettrine} % The lettrine is the first enlarged letter at the beginning of the text

\usepackage{enumitem} % Customized lists
\setlist[itemize]{noitemsep} % Make itemize lists more compact

\usepackage{abstract} % Allows abstract customization
\usepackage{titlesec} % Allows customization of titles
\renewcommand\thesection{\Roman{section}} % Roman numerals for the sections
\renewcommand\thesubsection{\roman{subsection}} % roman numerals for subsections
\titleformat{\section}[block]{\large\scshape\centering}{\thesection.}{1em}{} % Change the look of the section titles
\titleformat{\subsection}[block]{\large}{\thesubsection.}{1em}{} % Change the look of the section titles

\usepackage{fancyhdr} % Headers and footers
\usepackage{titling} % Customizing the title section

\usepackage{hyperref} % For hyperlinks in the PDF

\usepackage{graphicx} %package to manage images
\usepackage{amsmath}

\usepackage{pdflscape}

\pretitle{\begin{center}\Huge\bfseries} % Article title formatting
\posttitle{\end{center}} % Article title closing formatting
\pdfoutput=1
\title{Analytical modelling of temperature effects on synapses} % Article title
\author{
	\textsc{Dominik S. Kufel}\thanks{dominic.kufel@gmail.com - to who correspondence should be addresed} \\[1ex] % Your name
	\normalsize The Polish Children's Fund, Pasteura 5a, 02-093 Warsaw, Poland\\ % Your institution
	\normalsize S. Staszic High School no. 1, Raclawickie 26, 20-043 Lublin, Poland\\ % Your institution
	\and % Uncomment if 2 authors are required, duplicate these 4 lines if more
	\textsc{Grzegorz M. Wojcik}\thanks{gmwojcik@umcs.pl} \\[1ex] % Second author's name
	\normalsize Maria Curie-Sklodowska University in Lublin, Akademicka 9, 20-033 Lublin, Poland \\ % Second author's institution
}
\date{\vspace{-2ex}}
\renewcommand{\maketitlehookd}{%
\begin{abstract}
It was previously reported, that temperature may significantly influence neural dynamics on different levels of brain modelling. Due to this fact, while creating the model in computational neuroscience we would like to make it scalable for wide-range of various brain temperatures. However currently, because of a lack of experimental data and an absence of analytical model describing temperature influence on synapses, it is not possible to include temperature effects on multi-neuron modelling level.

In this paper, we propose first step to deal with this problem: new analytical model of AMPA-type synaptic conductance, which is able to include temperature effects in low-frequency stimulations. It was constructed on basis of Markov model description of AMPA receptor kinetics and few simplifications motivated both experimentally and from Monte Carlo simulation of synaptic transmission. The model may be used for efficient and accurate implementation of temperature effects on AMPA receptor conductance in large scale neural network simulations. This in fact, opens wide-range of new possibilities for researching an influence of temperature on brain functioning.
\end{abstract}
}

\begin{document}

% Print the title
\maketitle

%----------------------------------------------------------------------------------------
%	ARTICLE CONTENTS
%----------------------------------------------------------------------------------------

\section{Introduction}

From medical perspective, it was suggested that tight control of brain temperature in patients suffering posttraumatic period is highly recommended \cite{medicalbullshit:dg}. However, direct mechanism of influence of temperature on neural dynamics is still uncertain and better understanding of temperature effects on different levels of brain functioning may be useful in developing sophisticated treatment methods of different neurological disorders associated with increased or decreased brain temperatures \cite{dietrich:dg}. 

On level of single neurons, different effects of temperature were observed. The most important from perspective of neural dynamics are: 
1) Temperature influences membrane resting potential (\cite{hodgkinandhuxley:dg}, \cite{buzatu:dg}).
2) Temperature affects ion channels dynamics (\cite{hille:dg}, \cite{steratt:dg}). 
3) Temperature affects synaptic transmission (\cite{asztely:dg}, \cite{weight:dg}, \cite{schiff:dg}). 

However, temperature effects on membrane resting potentials (Goldman-Hodgkin-Katz equation) and ion channels dynamics \cite{hodgkinandhuxley:dg} are now well-known, an influence of temperature on synaptic transmission was proven to be more elaborating than on voltage-gated ion channels (see: \cite{roth:dg}). This may come from the fact, that for the various processes involved in synaptic transmission, like presynaptic release of neurotransmitter, dynamics of a vesicle pore, diffusion of neurotransmitter, binding of the neurotransmitter, kinetics of postsynaptic receptors, temperature influence is different and may significantly modify overall effect of heat on synaptic transmission \cite{fuxe:dg}.

Generally, in creation of models in computational neuroscience it is useful to make it easily scalable for different temperatures. This is especially important, due to the fact, that some of the neurobiological experiments are conducted in lower temperatures (for example \textit{in vitro} studies). Actually, a better knowledge about temperature dependence of synaptic transmission is a basis of linking \textit{in vitro} and \textit{in vivo} studies. Furthermore, optimal way of including full description of temperature effects in neural simulations may open computational brain research for new thermodynamic arguments. 

We may tackle the problem of including temperature effects on synapses from different perspectives. 

(1) A first approach is to include some coefficients associated with temperature in phenomenological synapse models. Through time, different, phenomenological methods of modelling of synaptic conductance were developed (alpha function, dual-exponential functions, single exponential function etc.). In practice, to include effects of temperature we would have to multiply all of the time constants and amplitudes of phenomenological functions by some (probably different) factors associated with temperature. However, there are different problems related to this approach. Firstly, we do not know the values of the temperature coefficients, by which we would like to multiply parameters of function (which are dependent on synaptic conductance function we use). This values would have to be obtained from extra experiment in different temperature, fitted directly to same modelling function. Furthermore, due to the fact that we do not know, whether all of the coefficients scale linearly with temperature, for each temperature we would have to perform additional experiment (which is not always possible for ex. when we are conducting \textit{in vitro} research).

(2) A second approach to take temperature effects into account is to model synapses on microphysiological level - so to investigate temperature effects on kinetics of synaptic receptors (proteins, which conformation is described by kinetic scheme). To include temperature effects we multiply all of the kinetic rate constants between different conformational states of protein by some coefficients dependent on temperature. This approach was previously taken experimentally (\cite{postlethwaite:dg},\cite{cais:dg}). However, the problem is that all of the temperature coefficients, which scale rate constants, are specific for given kinetic scheme. So, even if we have found temperature coefficients in one kinetic scheme, they are invalid for others (unless, we find some way to link different kinetic schemes, which is still not possible apart from linking very simple kinetic models \cite{linkingkinetics:dg}). So again, for our model we need to additionally find all of the temperature coefficients each time, which may be elaborating task for some types of synapses. 

Actually, both of the approaches have the same essence. In second approach, to include temperature effects we multiply all of the kinetic rates by some temperature factors. In the first approach we multiply amplitude and time constants of functions. However, amplitude and time constants in phenomenological modelling, under certain assumptions, may be interpreted as a combination of different kinetic rates, as it was proposed before \cite{destexhe:dg}. 

Generally, the problem of including temperature effects on synapse modelling is complex. Both first and second approaches are hard to generalize for different phenomenological functions describing synaptic conductance or different microphysiological kinetic schemes and need additional experiments to conduct, which prevents previously developed model to be easily scalable for wide-range of temperatures. 

In this paper, novel approach to problem of including temperature effects on modelling synapses was proposed. On basis of previous experimental and numerical research we construct assumptions of new analytical model to include temperature effects in modelling of $\alpha$-amino-3-hydroxy-5-methyl-4-isoxazolepropionic acid (AMPA) receptor. Firstly, we propose simplifications of experimental kinetic scheme (by \cite{postlethwaite:dg}) to allow for analytical solution of problem. Secondly using Monte Carlo simulation and Markov modelling we introduce concept of uncoupling of differential equation system describing AMPA receptor kinetics. Thirdly, after solving set of differential equations we compare results of constructed model with numerical and experimental data. Finally, we suggest that our model is capable to include temperature effects in neural dynamics simulations at low frequencies, regardless to phenomenological function of AMPA synaptic conductance used. This in fact allows, for the first time, to optimally and accurately simulate neural dynamics in different temperatures, without performing any additional experiments. 
%------------------------------------------------

\section{Methods and model}

\paragraph{Monte Carlo simulation of synaptic transmission}
For additional information to prove assumptions and validate some of the results of the analytical model, Monte Carlo simulation of synaptic transmission was used. Monte Carlo simulation was constructed on basis of assumptions and parameters by \cite{postlethwaite:dg} - code in MCell simulator \cite{mcell:dg} of their original simulation is available \footnote{\scriptsize{https://senselab.med.yale.edu/modeldb/ \newline showModel.cshtml?model=85981}}. To account for proper average responses, we assumed variable vesicle location above central postsynaptic density(PSD). Additionally, four neighbouring PSDs were included, positioned in the same geometry as used in model by \cite{postlethwaite:dg}. 

\begin{figure}[width=.7\textwidth]
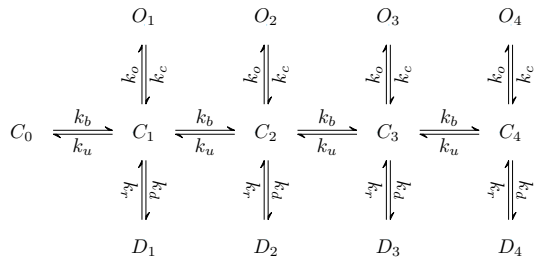

\caption{\scriptsize{Scheme 1 - modified kinetic scheme model by \cite{postlethwaite:dg},independent binding was assumed (see below) and no transitions between desensitized states (with minor influence on accuracy of results).}} 
\scalebox{.75}{
\schemestart
 $C_0$
 \arrow(A--B){<=>[$k_b$][$k_u$]}
$C_1$
 \arrow(@B--Y){<=>[$k_o$][$k_c$]}[90.] $O_1$
 \arrow(@B--Z){<=>[$k_d$][$k_r$]}[-90.] $D_1$
 \arrow(@B--C){<=>[$k_b$][$k_u$]}[0.] $C_2$
 \arrow(@C--YY){<=>[$k_o$][$k_c$]}[90.] $O_2$
 \arrow(@C--ZZ){<=>[$k_d$][$k_r$]}[-90.] $D_2$
 \arrow(@C--D){<=>[$k_b$][$k_u$]}[0.] $C_3$
 \arrow(@D--YY){<=>[$k_o$][$k_c$]}[90.] $O_3$
 \arrow(@D--ZZ){<=>[$k_d$][$k_r$]}[-90.] $D_3$
 \arrow(@D--E){<=>[$k_b$][$k_u$]}[0.] $C_4$
 \arrow(@E--YYYY){<=>[$k_o$][$k_c$]}[90.] $O_4$
 \arrow(@E--ZZZZ){<=>[$k_d$][$k_r$]}[-90.] $D_4$
\schemestop
}
\end{figure}

\paragraph{Analytical model}

Our model is based on the following numerical and experimental findings:

\textbf{(1)} Acceleration in postsynaptic AMPA receptor kinetics is the predominant effect of temperature on altered synaptic responses \cite{postlethwaite:dg} at low frequencies (to avoid effects of short-term synaptic plasticity). This assumption leads to the fact, that in our model the problem of temperature effects on synapses was simplified by considering temperature effects only on AMPA receptor kinetics, rather than also on modified presynaptic release and/or neurotransmitter diffusion dynamics. Furthermore, we assumed that to include temperature effects on AMPA receptor kinetics it is enough to multiply all of the rate constants by single temperature coefficient $Q_{10}$. \footnote{This approach is analogical to way of including temperature effects on voltage-gated ion-channels \cite{hodgkinandhuxley:dg} and motivated by Arrhenius equation. $Q_{10}$ informs how many times given speed of reaction increases with $10 ^{\circ}C$ increase of temperature.}

\textbf{(2)} All of the state transitions except transition from closed to bound states in AMPA receptor kinetics (considering single mesh of Scheme 2) were assumed to be Markov models: time and voltage independent and dependent only on the occupancy of neighbouring states as was previously proposed by \cite{destexhe:dg}.

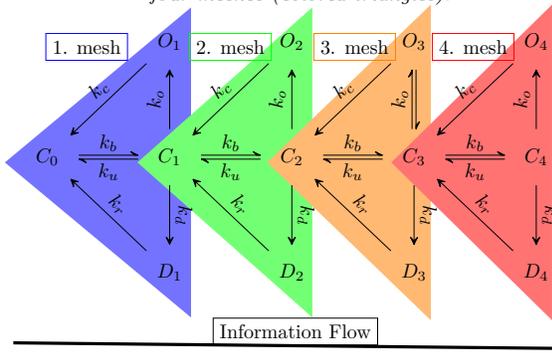
\begin{figure}[width=.8\textwidth]
\caption{\scriptsize{Scheme 2. Kinetic scheme used for construction of an analytical model consists four orders of subconductance(index numbers of states - we do not consider 0-th order) and four meshes (colored triangles).} }
\centering
\scalebox{.75}{
\begin{tikzpicture}[help lines/.style={thick,draw=black!50},node distance=2cm]

\draw[ultra thick, ->] (-0.4,-3.2) -- (9.4,-3.3);

\node[draw] at (4.6,-3.0) {Information Flow};

\fill[blue!55!white]  (-0.55,0) -- (2.75,-2.75) -- (2.75,2.75) -- cycle;
\fill[green!55!white]  (1.8,0) -- (4.9,-2.75) -- (4.9,2.75) -- cycle;
\fill[orange!55!white]  (4.1,0) -- (7.0,-2.8) -- (7.0,2.8) -- cycle;
\fill[red!55!white]  (6.3,0) -- (9.15,-2.8) -- (9.15,2.8) -- cycle;
\node[draw=blue!] at (0.9,2.05) {1. mesh};
\node[draw=green!] at (3.45,2.05) {2. mesh};
\node[draw=orange!] at (5.65,2.05) {3. mesh};
\node[draw=red!] at (7.75,2.05) {4. mesh};

\schemestart
 $C_0$
 \arrow(A--B){<=>[$k_b$][$k_u$]}
$C_1$
 \arrow(@B--Y){->[$k_o$]}[90.] $O_1$
 \arrow(@Y--@A){->[$k_c$]}[-30.]
 \arrow(@B--Z){->[$k_d$]}[-90.] $D_1$
 \arrow(@Z--@A){->[$k_r$]}[30.]
 %%%%%%%%%%%%%%%%%%%%%%%%%%%%%%
 \arrow(@B--C){<=>[$k_b$][$k_u$]}[0.] $C_2$
 \arrow(@C--YY){->[$k_o$]}[90.] $O_2$
 \arrow(@YY--@B){->[$k_c$]}[-30.]
 \arrow(@C--ZZ){->[$k_d$]}[-90.] $D_2$
 \arrow(@ZZ--@B){->[$k_r$]}[30.]
 %%%%%%%%%%%%%%%%%%%%%%%%%%%%%
 \arrow(@C--D){<=>[$k_b$][$k_u$]}[0.] $C_3$
 \arrow(@D--YYY){<=>[$k_o$]}[90.] $O_3$
 \arrow(@YYY--@C){->[$k_c$]}[-30.]
 \arrow(@D--ZZZ){->[$k_d$]}[-90.] $D_3$
 \arrow(@ZZZ--@C){->[$k_r$]}[30.]
 %%%%%%%%%%%%%%%%%%%%%%%%%%%%%%%%%
 \arrow(@D--E){<=>[$k_b$][$k_u$]}[0.] $C_4$
 \arrow(@E--YYYY){->[$k_o$]}[90.] $O_4$
 \arrow(@YYYY--@D){->[$k_c$]}[-30.]
 \arrow(@E--ZZZZ){->[$k_d$]}[-90.] $D_4$
 \arrow(@ZZZZ--@D){->[$k_r$]}[30.]
\schemestop
\end{tikzpicture}
}
\end{figure}

\vspace{5mm} %5mm vertical space

\textbf{(3)} As it was suggested by \cite{postlethwaite:dg} temperature effects are mediated by driving AMPARs to higher subconductance states. To include higher subconductance states in analytical model of AMPA receptor few simplifications of complex 13 states and 30 transitions kinetic scheme by \cite{postlethwaite:dg} (Scheme 1) were proposed. Generally, Scheme 1 was re-written into simplified form of Scheme 2. This form uses symmetry of states and transitions in Scheme 1. This symmetry comes from the fact, that (when combined with (6)) we may divide kinetic scheme (Scheme 1) to overall four orders of subconducting states. 

Additionally, one further analytical simplification is made: modification in directions of transitions with rates $k_r$ and $k_c$ where taken to 'relieve' state $C_i$ tangled in many transitions with minor influence on solutions for open and desensitized states of AMPA receptor. This approach allows for a much easier analytical solution of differential equations describing kinetic scheme.

\begin{figure}[width=1.5\textwidth]
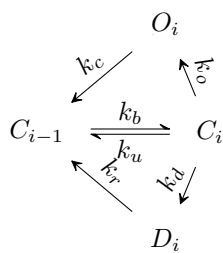

\caption{\scriptsize{Scheme 3. Single mesh described by an independent pair of coupled differential equations}}
\scalebox{1.0}{
\schemestart
$C_{i-1}$
 \arrow(A--B){<=>[$k_b$][$k_u$]}
$C_{i}$
 \arrow(@A--D1){<-[$k_r$]}[-40.] $D_{i}$
 \arrow(@A--Y1){<-[$k_c$]}[40.] $O_{i}$
 \arrow(@B--@D1){->[$k_d$]}[-90.] 
 \arrow(@B--@Y1){->[$k_o$]}[90.] 
\schemestop
}
\end{figure}

\textbf{(4)} Separately, the 1st order (and symmetrically, with additional assumptions described below, i-th order) of AMPAR kinetic states was assumed to behave as a single mesh, according to Scheme 3 (here, for $i=1$). The form $C_0$ was considered to be in excess compared to other states in this scheme. This is directly the case, when very few receptors bind glutamate, so that nearly all receptors remain in form $C_0$. This is motivated by comparing number of channels in different states in Monte Carlo simulation(see Figure 4). Therefore fraction of channels in $C_0$ state (and consequently due to further simplification in each $C_{i-1}$ state in i-th mesh) is considered always as 1. This assumption is taken because considering fractions of channels in all other states multiplied by some function dependent on time (see assumption (6)) leads to much less trivial system of differential equations with much more complexed solution. Furthermore, without this assumption including higher subconductance states would be not possible in an analytical way (see below).

After generalization (assumption (5)) for i-th order, the assumption is true for every mesh of kinetic scheme (see below and Scheme 2). 

\textbf{(5)} To account for higher subconductance states a concept of \textit{uncoupling of a set of differential equations} introduced below was used. 

In general, this what influences complexity of system of differential equations describing kinetic states are multiple transitions between some state and the others. To find functions of AMPA receptor open \footnote{experimental data suggest, that only open state of protein can influence synaptic conductance \cite{smith:dg}} states ($O_1$,$O_2$,$O_3$,$O_4$) dependent on time in given kinetic state with $m$ states,one would have to write $m-1$ coupled differential equations, which complexity is proportional to number of transitions between different states \cite{destexhe:dg}. However, generally for some cases system of coupled differential equations may be uncoupled with minor loss of accuracy of solution. In our problem, we would like to uncouple i-th order (where $i$ $\epsilon$ <1;4>) of bound state $C_i$ with (i+1)-th order of bound state $C_{i+1}$ (which are only causes of coupling between different orders of kinetic scheme). 

Our approach is based on an idea, taken as an analogy from a very simple economical model. Consider an agent selling some product on the market. Now, let's assume that decision of an agent about selling the product is dependent on current state of a market (whether its condition is good or bad etc.). So, in general market is affecting seller by its current state and seller is affecting market by selling product and therefore changing its state. Then, an information (about decisions of agent or state of market) goes both directions: from agent to market and from market to agent. This situation may be described by differential equation. Sometimes, the form of differential equation may not be trivial. However, to simplify this situation we use an assumption, that if market is huge, an influence of a single agent on a market will not be significant. Then, the essence of this simplification is the fact, that information flows only from market to agent (rather than both directions), therefore leading to much easier mathematical description of this problem. 

In case of kinetic scheme simplification, a similar approach was used: it is enough to assume, that transitions from higher order subconductance states to lower order have little (no) influence on lower states. It means, that from perspective of lower order of conductance, all of the transitions from higher order states are neglectable. However, from perspective of higher order of conductance, all of the transitions from lower order are still significant (see information flow on Scheme 2).

However, validity of this assumption requires, that sum of all of the transition rates from higher to current order of subconductance has to be much smaller than transition rate from current order to higher plus from lower to current order (see: Scheme 2). This assures, that relatively little information goes from higher to lower subconductance orders, so we may neglect this influence of higher states in construction of the model. 

In mathematical terms, assumption of uncoupling for i-th order may be written as:

\begin{equation}
\label{eq:assumption}
k_b(t) (x_{i-1}+x_i) >> k_c y_{i+1} + k_r z_{i+1} + k_u (x_{i+1}+x_i)
\end{equation}

Particulary, for 1st order we  get:
\begin{equation}
\label{eq:assumption-example}
k_b(t) (x_0 + x_1) >> k_c y_2 + k_r z_2 + k_u (x_2+x_1)
\end{equation} 

This method allows us to uncouple set of twelve coupled differential equations with complex formulation to set of 4 pairs of differential equations (coupled only in pairs, rather than between different orders of subconductance). Thanks to this approach, we are able to include higher order subconductance states and, as a result, find an analytical solution of the problem. 

Actually, this approach is similar to assumption (4): we may say that from perspective of (i+1)-th order subconductance state, the fraction of channels in i-th order subconductance bound state ($C_i$) is perceived as $1$. However, to differentiate how huge this $1$ is absolutely, for each i-th order of conductance $\lambda_i (t)$ function is introduced. 
$\lambda_i (t)$ scales 'relative fraction of channels in each state' to 'absolute (scaled samely to all of the orders of kinetic scheme) fraction'.

\textbf{(6)} Glutamate binding was assumed to be independent, similary to model by \cite{robertandhowe:dg} This departure from \cite{postlethwaite:dg} was motivated by a disproportionate increase of complexity of analytical solution of a set of differential equations when assuming cooperative binding. Cooperative binding requires a more elaborating form of kinetic scheme, mostly due to assumptions associated with including AMPAR higher order subconductance states taken in this model. \\
 
\textbf{(7)} Glutamate concentration was assumed to be time-dependent according to single exponential decay function, which parameters were fitted to data of glutamate concentration at synaptic cleft (and consequently at PSD) in Monte Carlo model of synaptic transmission. Therefore function of binding rate constant from closed to bound states has a form dependent on concentration of glutamate at PSD:

\begin{equation}
\label{eq:emc}
k_b(t)=k_b A e^{-\omega t}
\end{equation}

where parameters $\omega$ [1/s] and $A$ [molar] were fitted from averaging of glutamate concentration (in cleft) in Monte Carlo simulation of synaptic transmission.  
This approach was taken due to several causes. Departure from model based on neurotransmitter concentration occuring as a pulse (described by Dirac Delta at $t_{pulse}$) proposed by \cite{destexhe:dg} or \cite{destexhe1994:dg} was motivated by availability of direct data of glutamate concentration on PSDs in Monte Carlo simulation and unit inconsistency problem\footnote{Modelling of glutamate concentration at PSD as a pulse of transmitter (motivated by the fact, that concentration of neurotransmitter rises and falls rapidly \cite{coloqhun:dg}) in $t=t_{pulse}$ and assumption about considering fraction of channels in $C_0$ state as 1 leads to unconsistency in units in function of fraction of channels in open state versus time (just as in derivation of alpha function in Appendix B of \cite{destexhe:dg}) and therefore not being able to normalize to 1, which would make model unable to reproduce data quantitatively. However, one has to admit that modelling of neurotransmitter concentration by Dirac Delta has some advantages (see \cite{destexhe1994:dg} and \cite{destexhe:dg}) and may be useful in obtaining dual-exponential solution for synaptic conductance, which can be easily interpreted \cite{roth:dg}.}. Furthermore, an influence of temperature on glutamate concentration was also investigated (to investigate whether diffusion of neurotransmitters altered by temperature may lead to more accurate solution in comparison to experimental data). 

Dual-exponential function was also considered as a fitting method of glutamate concentration, but generally the problem was that solution of system of differential equations describing kinetic scheme were getting extremely complex (and therefore AMPAR higher order subconductance states could not be included), unproportionally to gain in accuracy of analytical model. 

\textbf{(8)} According to experimental data, AMPA receptors are tetramers \cite{rosenmund:dg}. Conductance of AMPA receptor can be described as a sum of conductances of all orders of subconductance multiplied by different constants for different orders of states in kinetic scheme (as suggested before \cite{sahara:dg}):
\begin{equation}
\label{eq:sum}
g(t) = g_4 \sum_{i=1}^{n=4} a_i y_i(t) 
\end{equation}
where, $g_4$ is a peak conductance of a channel in 4-fold bound state, $n$ is a number of orders in kinetic scheme and $a_i$, $y_i$ are scalling factor and fraction of open channels in the i-th state respectively. \\
Conductances of different orders of states in kinetic scheme were set as fractions of the peak conductance at the 4-fold bound state $O_4$ ($O_1$ : $a_1=0.1$, $O_2$ : $a_2=0.4$, $O_3$ : $a_3=0.7$, $O_4$ : $a_4=1.0$) as it was proposed by \cite{postlethwaite:dg} and motivated by previous experimental work by \cite{smith:dg}. This assumption in turn, suggests, that we may break down the problem of finding conductance into sum of four functions, which suggest that we should operate on so called before 'orders' of kinetic scheme. 

\begin{figure}[h]
\caption{\scriptsize{Fraction of AMPAR channels in unbound state $C_0$. Fraction of channels decays from $1$ to about $0.8$ in $3 ms$. It shows, why we assumed that fraction of channels in state $C_{i-1}$ in first mesh is in excess when compared to the other states.}}
\centering
\includegraphics[width=0.5\textwidth]{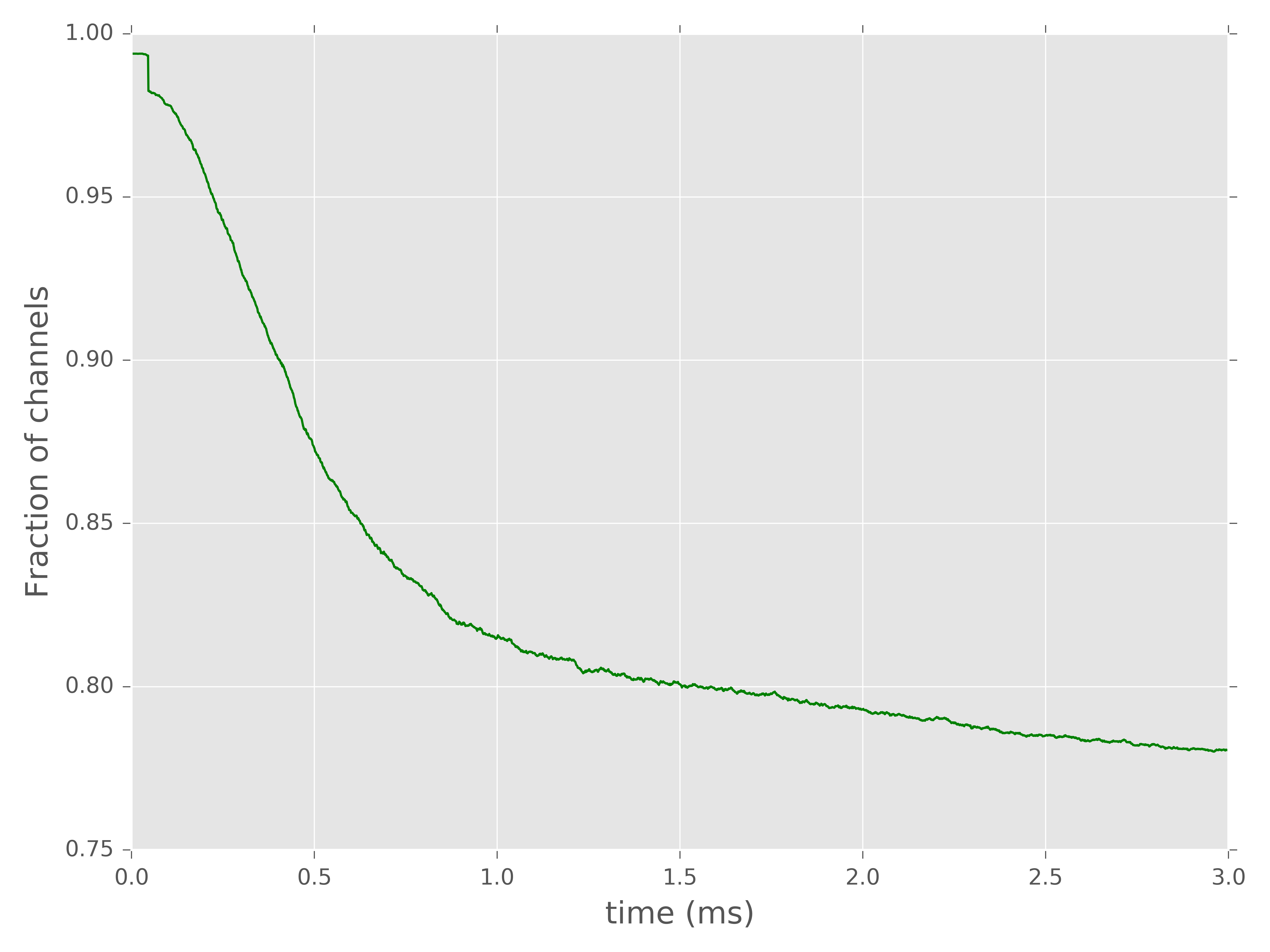}
\end{figure}

Combination of all of assumptions made before, simplifies complex kinetic scheme, containing 13 states and multiple transitions (described by 12 coupled differential equations). 
Modification of directions of transitions (from $O_i$ and $D_i$) with uncoupling concept splits system of differential equations to four, one-side dependent and specular between themselves, meshes of kinetic scheme. Furthermore, by assuming fraction of channels in state $C_{i-1}$ for i-th order of scheme to be equal to 1 we are able to solve coupled pair of differential equations in every single mesh.
Thus, simplification leads to 4 independent pairs of coupled differential equations, which solution for first order of conductance we know explicitly and for higher orders we use $\lambda_i (t) = x_{i-1} (t)$ (which is solution in respect to fraction of channels in state $C_{i-1}$ of pair of differential equation for (i-1)-th order), which assures information flow from lower to higher orders of AMPAR subconductance. 

Making these assumptions, one obtains general system of differential equations, describing i-th order of kinetic states (Scheme 2):

\begin{equation}
\label{eq:firstdifferential}
\frac{dx_{i}}{dt}=k_b A e^{-\omega t} \lambda_i (t) - (k_o+k_u+k_d)x_{i}(t) 
\end{equation}

\begin{equation}
\label{eq:seconddifferential}
\frac{dy_{i}}{dt}=k_o x_{i}(t) - k_c y_{i}(t)
\end{equation}

where $y_i(t)=[O_{i}(t)]$ is a fraction of channels in an open state of i-th order, $x_{i}(t)=[C_{i}(t)]$ is a fraction of channels in a bound state of i-th order,  $\lambda_i (t)$ is a function to convert fraction of all channels to same absolute scale (not only relative for each order) - for i-th order of Scheme 2 it equals to solution with respect to fraction of channels in state $C_0$ of two differential equations of (i-1)-th order: $\lambda_i (t) = x_{i-1}(t)$. Thanks to this approach we may include higher subconductance states of AMPA receptor with analytical approach, due to the fact of uncoupling of differential equations describing kinetic scheme. 

The solution for the first order of coupled differential equations (5) and (6) with boundary conditions $y_1(0)=0$ and $x_1(0)=0$ is: 

\begin{equation}
\label{eq:solution}
\resizebox{1.\hsize}{!}{$y_1(t)=\frac{A k_b k_o}{S P} e^{-\omega t} + \frac{A k_b k_o}{R P} e^{-(P + \omega) t} 
- \frac{A k_b k_o}{R S} e^{-k_c t}$}
\end{equation}

where $S= k_c - \omega$, $P=k_d+k_o+k_u-\omega$, $R=-k_c+k_d+k_o+k_u$.

As it is possible to be seen, first-order approximating function is a sum of exponents (as suggested by \cite{destexhe:dg}).

The full solution for all of the orders of kinetic scheme (Scheme 2) can be found in Appendix A. 

\paragraph{Normalization of fraction of channels}

Due to the assumptions of model presented above, without any correction, it would not be able to fit to experimental data quantitatively (because sum of all of the fractions of channels in different states does not sum to one). The reason is the assumption (4) and (5): fraction of channels in state $C_i$ for (i+1)-th order mesh equals (relatively) to one. For example, fraction of channels in state $C_0$ for 1st order mesh equals to one. 

To fix this problem, we introduce normalization constant, which is an accuracy of assumption (4) with Monte Carlo simulation (accuracy is taken as $1-C_0$ in time of amplitude of function of fraction of open channels). Assuming, that most of the channels remain in state $C_0$ (Figure 4) we may see that this accuracy equals about 12.5\% (in a time of peak of synaptic conductance). So, to reproduce Monte Carlo results, we only have to multiply all of the fractions of channels in analytical model by 0.125. 

%------------------------------------------------

\section{Results}

\paragraph{Analytical model of AMPA receptor}

We found that, for kinetic rates fitted from model by \cite{postlethwaite:dg}, our model is able to reproduce both results of Monte Carlo simulation (describing fractions of AMPAR channels in different states: see Figure 2B by \cite{postlethwaite:dg}) and experimental data (describing shape of curve of AMPAR synaptic conductance: see \cite{postlethwaite:dg} Figure 1A). 

\paragraph{Number of channels}
Fractions of AMPAR channels in different states (Figure 5) are both qualitatively and quantitatively (after normalization: see methods) consistent with this obtained in Monte Carlo simulation (with mean 5\% error for function of open states in time). 

In the beginning, binding of glutamate leads AMPARs from unbound to bound states. The fraction of channels in bound state is dependent on glutamate concentration at PSDs and rate of unbinding in AMPA receptor kinetic scheme. Excluding for a while effects of unbinding, due to the fact, that the diffusion of neurotransmitters in synaptic cleft is considered as a random walk, mean distance of single neurotransmitter from location of release (vesicle pore) should increase proportionally to $\sqrt{N}$ where $N$ is number of steps (with certain displacement) in time. So, glutamate concentration should decay with $1/\sqrt{N}$ and number of bound states should increase proportionally to $1-\sqrt{N}$. However, because constantly, some of the AMPAR channels unbind neurotransmitters(transitioning from $C_1$ to $C_0$ state), function of bound states' fraction in time should be close to a 'flattened' $1-\sqrt{N}$.

Afterall, different conformations of AMPA receptor protein occur in time. The speed of conformational changes is proportional to temperature (reflected by scaling all of the rate constants of kinetic scheme by $Q_{10}$ parameter). 

Continous growth of fraction of channels in desensitized states comes from the fact that resensitizing rate of reaction is about three orders of magnitude smaller than rate of desensitization. Therefore, channels after entering, unlikely leave desensitized states and fraction of channels in desensitized states slowly approaches fraction of channels in all bound states.  

Generally, model is a little bit underestimating fraction of AMPAR channels in states bound and desensitized. This may come from an assumption in analytical model about directions of transitions between states from $O_i$ to $C_{i-1}$ (rather than $C_i$) and from $D_i$ to $C_{i-1}$ (rather than $C_{i-1}$) - in first order of subconductance states some of the transitions go back to unbound state (rather than first bound closed state as assumed in Scheme 1). Underestimation in fraction of channels in desensitized state comes from the fact, that due to modification in directions of transitions for first order of subconductance, smaller fraction of AMPARs is in bound state (resensitization due to its small transition likelihood has a minor influence).

\begin{figure}[h]
\caption{\scriptsize{Fraction of channels in different states as a function of time. Open scaled channels is a fraction of  channels in open states, multiplied by respective four order conductances.}}
\centering
\includegraphics[width=0.5\textwidth]{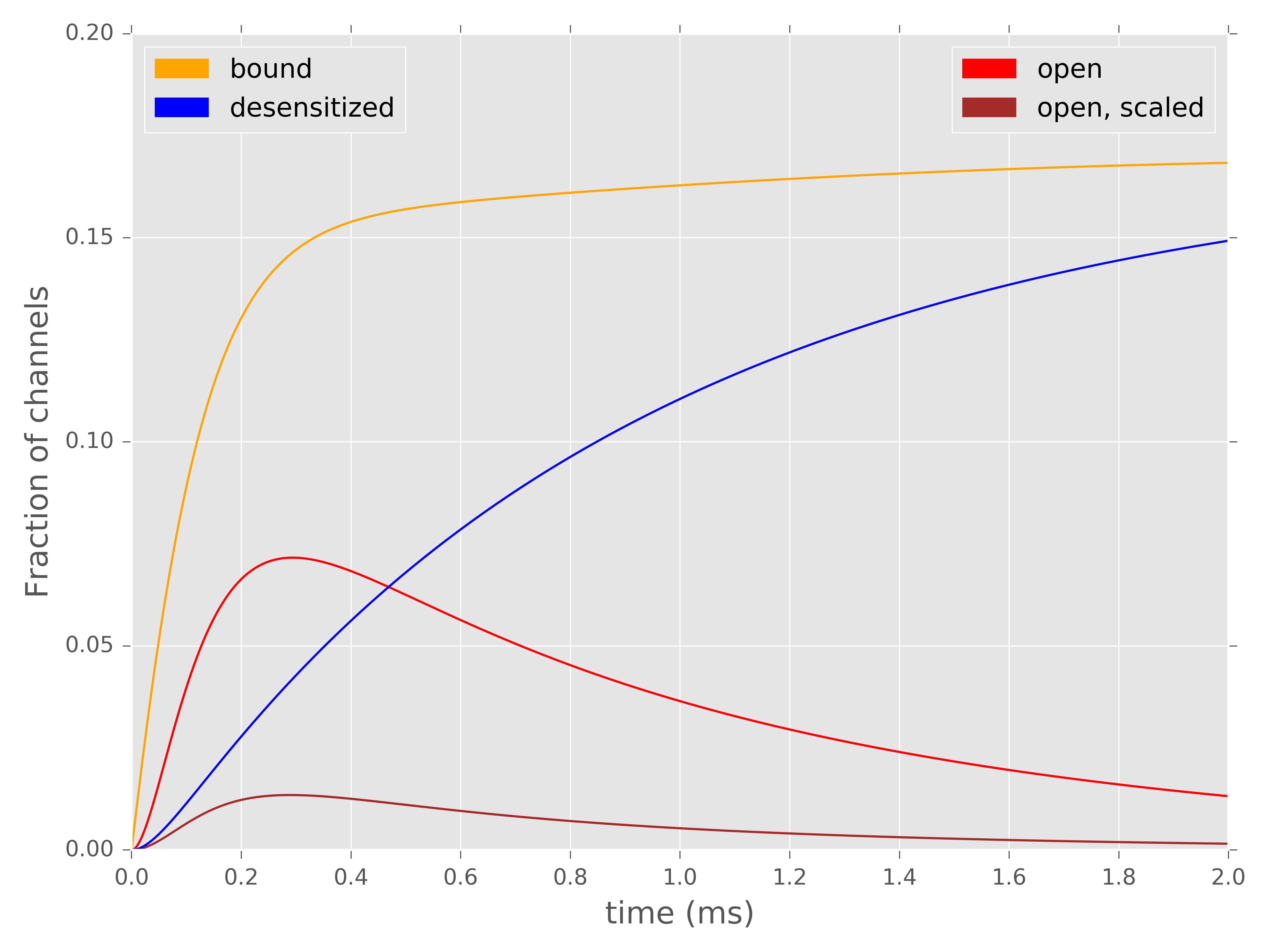}
\end{figure}

\begin{figure}[h]
\caption{\scriptsize{Conductance curve of AMPA synapse in $25 ^{\circ}C$ and $35 ^{\circ}C$. Conductance in $35 ^{\circ}C$ in comparison to $25 ^{\circ}C$ has bigger and faster peak.}}
\centering
\includegraphics[width=0.5\textwidth]{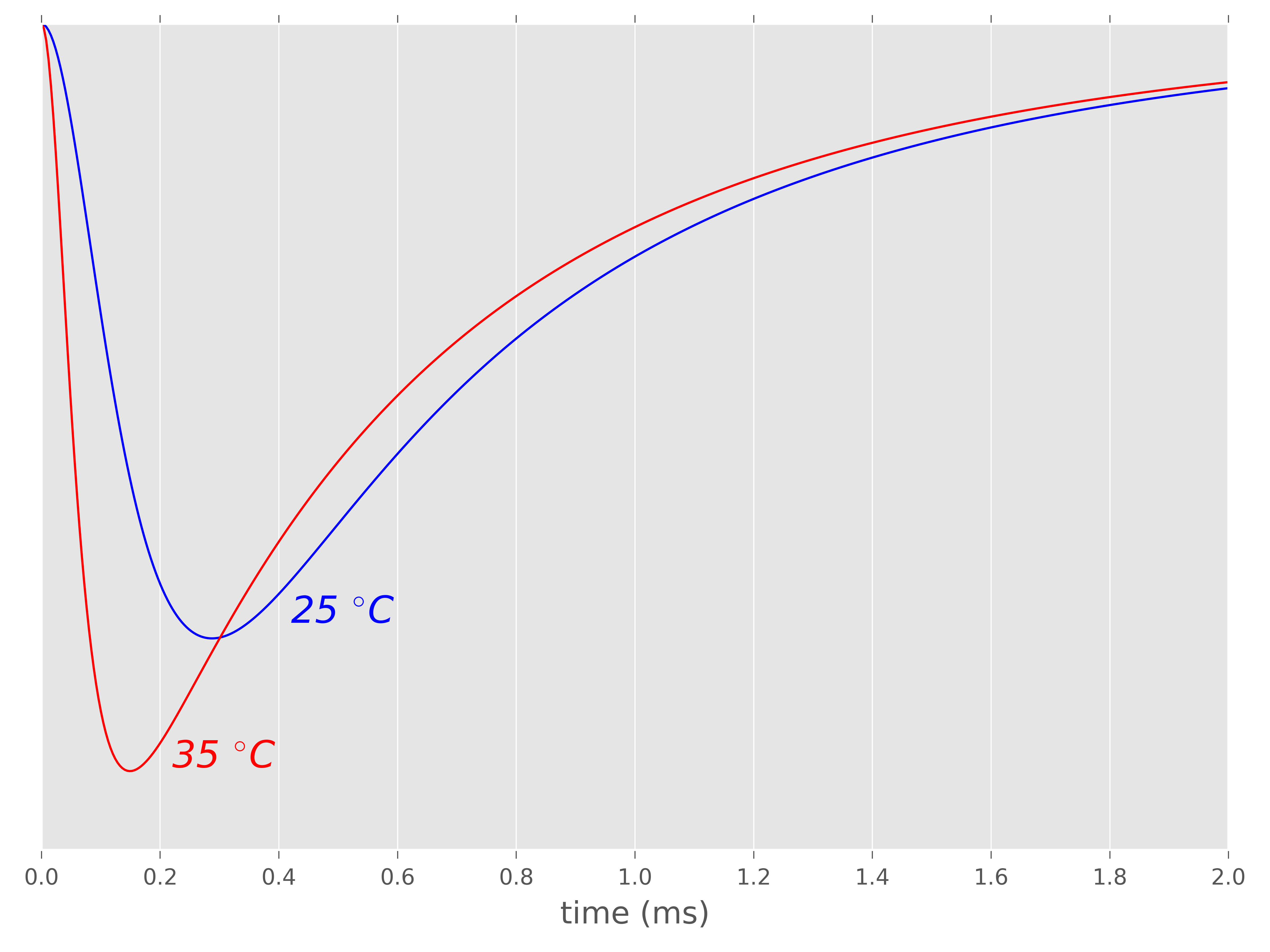}
\end{figure}

\paragraph{Synaptic conductance}

AMPAR conductance curve (Figure 6) obtained from analytical model is able to reproduce (with accuracy for relative amplitude and relative time of peak to < 5 \%, which is within the experimental error range) shape and scale of temperature effects on synaptic transmission (compare with Figure 1B by \cite{postlethwaite:dg}). In $35 ^{\circ}C$ both rise and decay time constant of synaptic conductance are smaller (faster). The peak conductance is bigger (for ratio about 1.25) and is reached faster in $35 ^{\circ}C$ in comparison to $25 ^{\circ}C$. 

However, analytical model predicts too rapid rise time of conductance in comparison to experimental data (see time of peak on Figure 6 and on Figure 1B by \cite{postlethwaite:dg}). This is due to the assumption about fitting function to glutamate concentration in synaptic cleft. In this model, only single exponential decay (from peak value at $t=0$) of glutamate concentration was assumed. It is only an approximation of reality, where evidently there is not infinitely fast rise of glutamate concentration (see Figure 7). Hence, strength of temperature effects on synaptic conductance is shaped by decay time constant of glutamate concentaration function. However, it is not possible to include any other, more elaborating forms (dual exponential function, fast and slow exponents etc.) of glutamate concentration function, due to no existence of closed analytical form of solution of differential equation system including higher subconductance states. 

\begin{figure}[h]
\caption{\scriptsize{Glutamate concentration obtained from Monte Carlo simulation in $25 ^{\circ}C$ and assumed fitting curve proportional to $ e^{- \omega t}$}}
\centering
\includegraphics[width=0.5\textwidth]{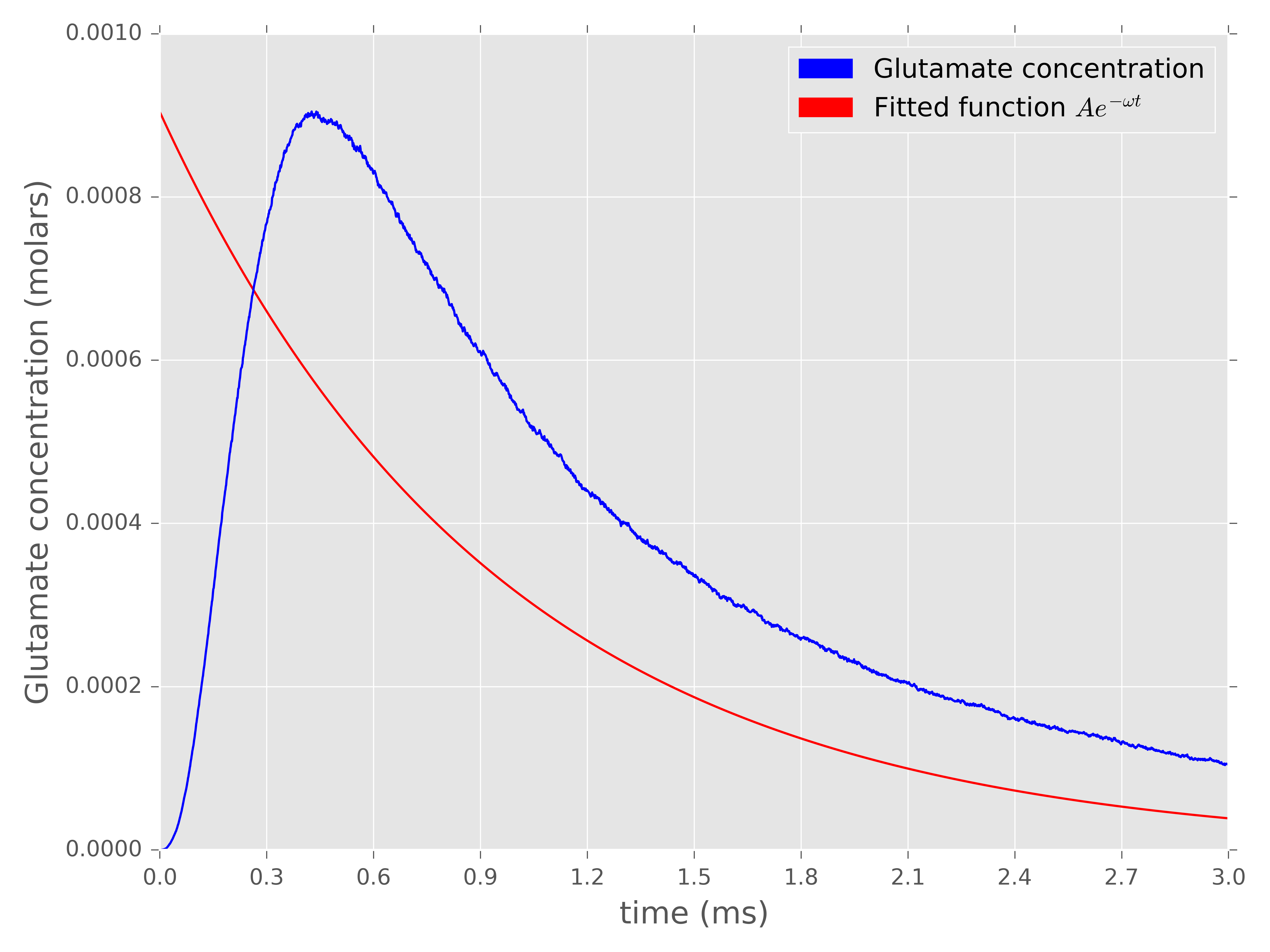}
\end{figure}

It was found that diffusion of neurotransmitters altered by diffusion cannot lead to more accurate solution in comparison to experimental data (Figure 8). Modifying $Q_{10_{diff}}$ coefficient of diffusion (effectively multiplying diffusion coefficient of glutamate) with temperature, leads to faster decay of glutamate concentration in synaptic cleft and faster time rise and decay of synaptic conductance (the physical meaning is, that glutamate molecules should simply move faster in higher temperatures). However, increasing $Q_{10_{diff}}$ causes also smaller AMPAR peak conductance, than this observed experimentally. 
This supports previous thesis by \cite{postlethwaite:dg} about predominant role of postsynaptic site of synapse in temperature effects on synapses. In turn, it may be important in the context of possible medical application: implementing drug, which is able to slow down receptor kinetics may lead to successful prevention of adverse temperature effects on dynamics of synapses (for example in state of hyperthermia or hypothermia in human brain). 

\begin{figure}[h]
\caption{\scriptsize{Influence of different $Q_{10}$ diffusion coefficients on synaptic conductance.}}
\centering
\includegraphics[width=0.5\textwidth]{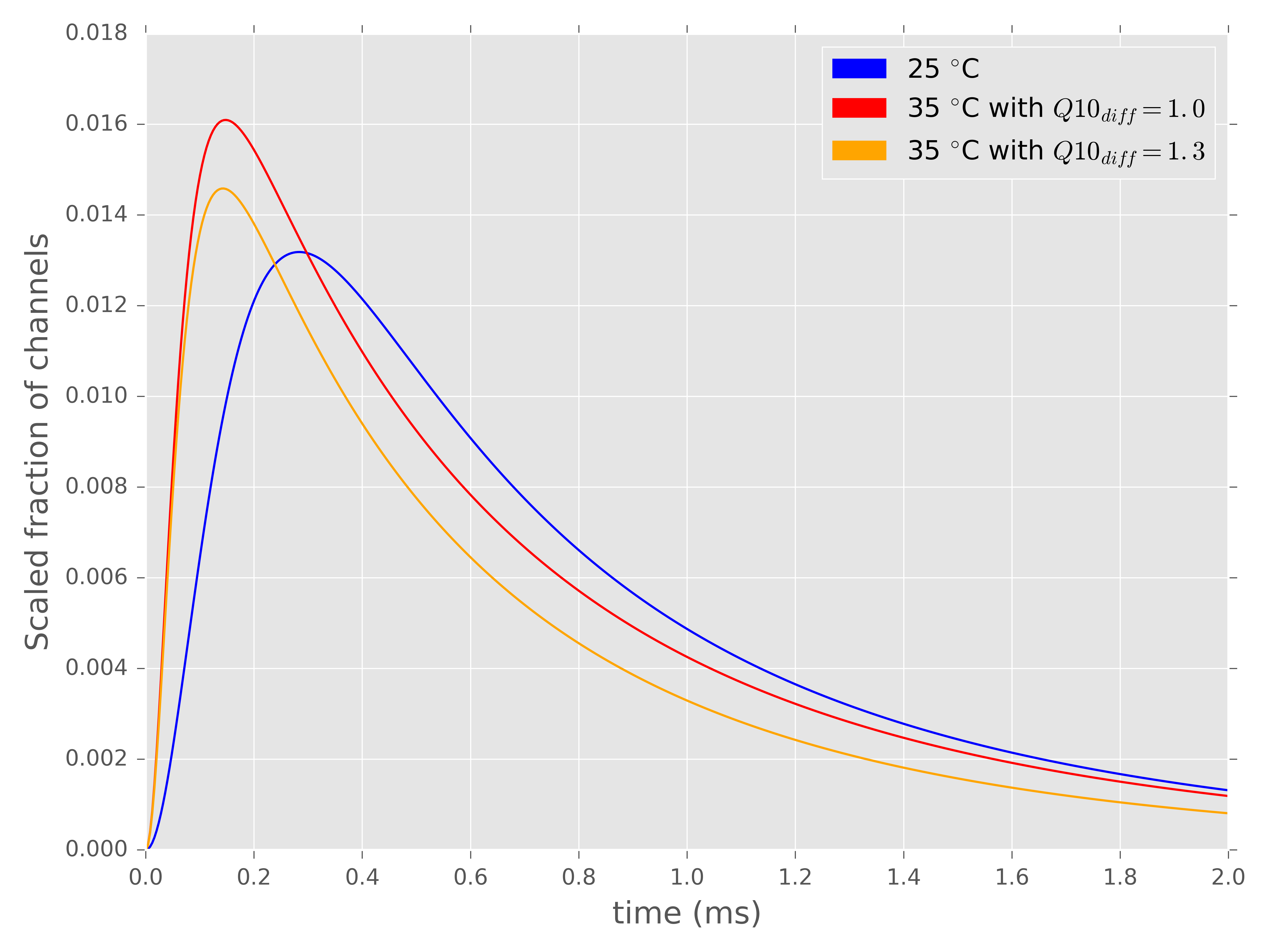}
\end{figure}

Significance of including higher subconductance states in analytical model was also investigated. It turned out, that including higher and higher states of subconductance leads to saturating increase of ratio (in $35 ^{\circ}C$ relatively to $25 ^{\circ}C$) of synaptic conductance peak amplitudes (for about 12\% in comparison to first order approximation), with minor influence on ratio of times of peak (Figure 9). Furthermore, it was found that it is possible to achieve same dynamics of AMPAR synaptic conductance (qualitatively and quantitatively) by using 3rd approximation (without including 4-th order) and changing fraction of the peak conductance at the 4-fold bound state for 3rd state from $0.7$ to $0.9$. Therefore, we support one of the work's results by \cite{postlethwaite:dg}, which claims that higher temperature leads AMPARs to higher conducting states(thus increasing conductance peak amplitude).

\begin{figure}[h]
\caption{\scriptsize{Different order approximations of higher subconductance states.}}
\centering
\includegraphics[width=0.5\textwidth]{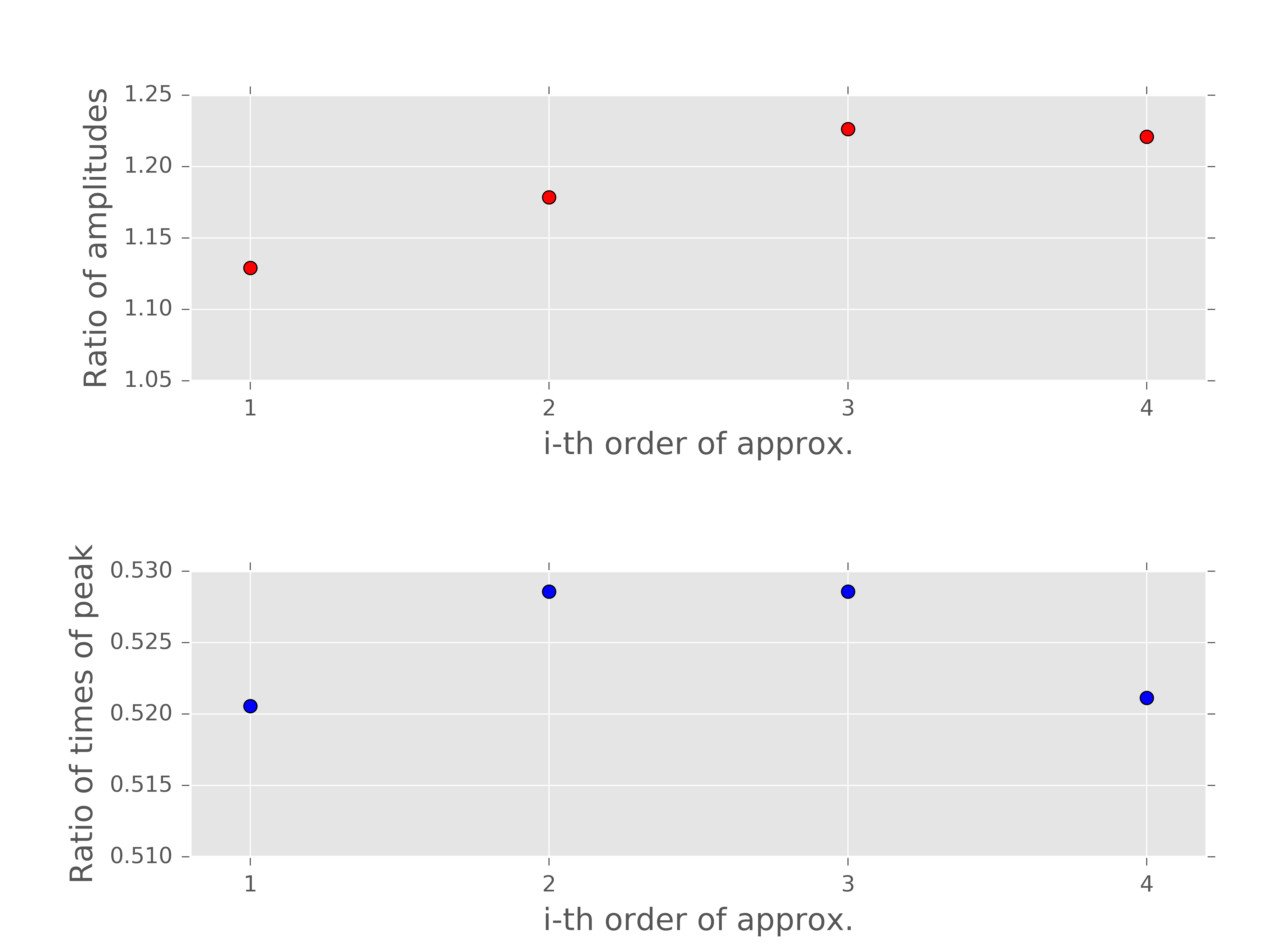}
\end{figure}

%------------------------------------------------

\section{Discussion}

\paragraph{Glutamate binding model}
An analytical model showed that in a light of assumptions we made independent glutamate binding is able to reproduce experimental and numerical results. However, considering complexity of biological systems and the role of variability in their behaviour, we do not argue, that independent binding is a biological reality, but rather a reliable approximation of system's average behaviour. In work by \cite{postlethwaite:dg} it was shown, that variability of both rise and decay time constants of AMPAR conductance as a function of the mEPSC peak amplitude was not successfully reproduced by an independent binding model. However, it is not possible to research this relationships using our model describing only an average behaviour. 

\paragraph{Temperature dependence of AMPA receptor conductance}
In this paper it is clearly showed, that increased temperature leads to bigger peak amplitude of AMPAR conductance, which is achieved faster than in lower temperature. However, this results are not easily interpretable in more general context - for example, an influence of modified synaptic conductance curve on temporal summation of signals across neuron's morphology. 

From the one point of view, due to the fact, that synaptic conductance has bigger peak amplitude in higher temperature, thus smaller number of EPSPs should elicit action potential than in lower temperature. From the other side, due to the fact that time constants of AMPAR conductance function decline faster, temporal summation should overlap less efficiently. So, this two naturally opposite features of AMPAR synaptic conductance hinder simple description of compounded temperature effects on multi-synaptic networks: it iss actually hard to strictly predict (because we do not know relative importance of amplitude and time constants of synaptic conductance curve) how it is going to influence temporal summation of synaptic signal. In the future, more detailed study on this topic may allow us to investigate temperature effects from level of single synapses to large neural-networks, which may help in better understanding of complex and paradoxical field interactions in brain imposed by temperature \cite{moser95:dg}. 

\paragraph{Uncoupling assumption accuracy}

To test an assumption of uncoupling of differential system equations we use Monte Carlo simulation of synaptic transmission for kinetic scheme proposed here (Scheme 3). This comes from the fact, that directions between some of the transitions were changed (see assumption (3)), therefore to capture correct dynamics of AMPAR conformational changes Monte Carlo simulation was employed.

For case of model we propose, differential equation system uncoupling assumption is fulfilled with different accuracy for every order (Figure 10). 

\begin{figure}[h]
\caption{\scriptsize{Uncoupling assumption validation for different orders of kinetic scheme. 'Left' and 'Right' are values of sides of equation (1) for parameters by \cite{postlethwaite:dg}.}}
\centering
\includegraphics[width=0.5\textwidth]{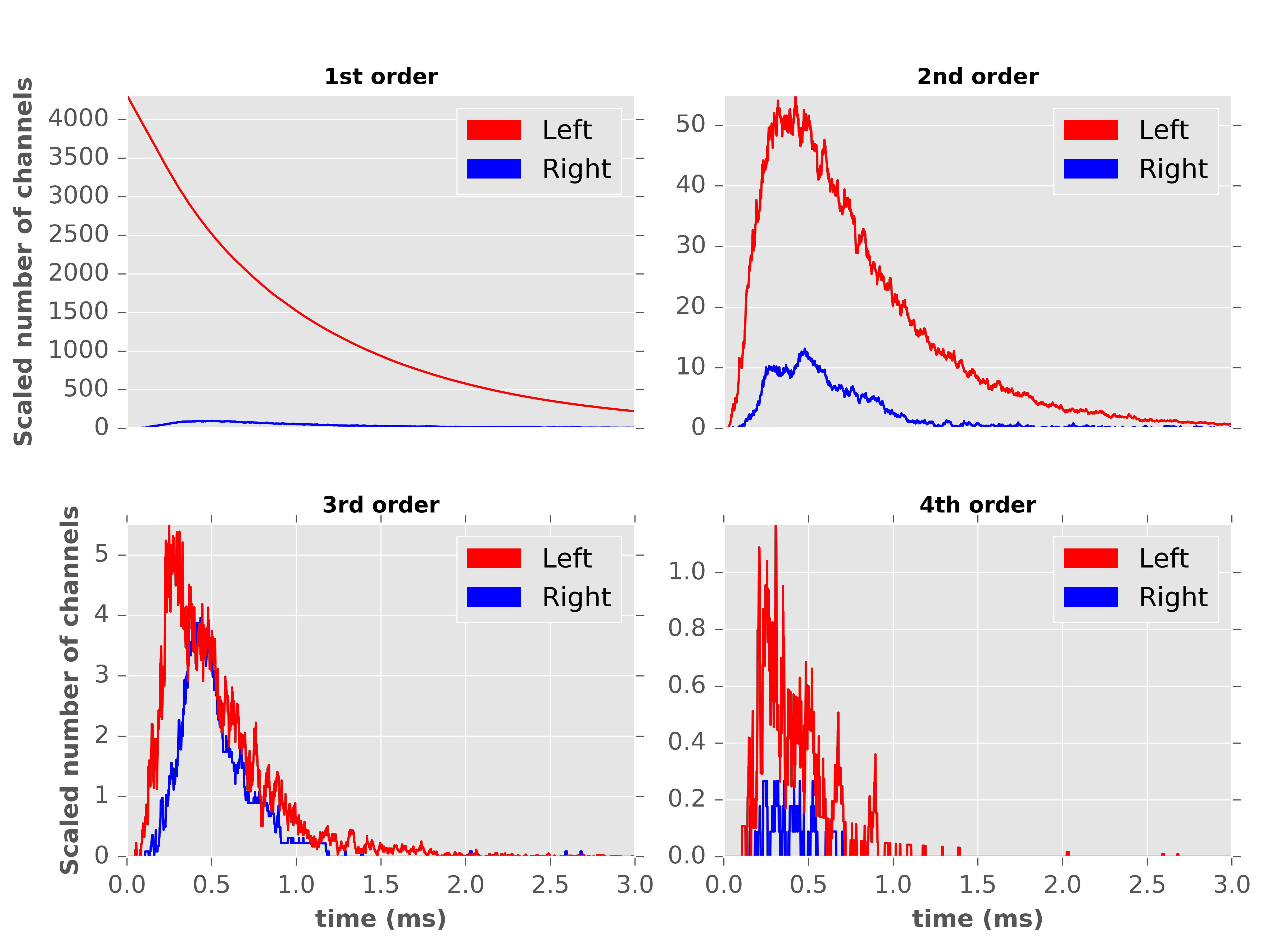}
\end{figure}

Generally, the assumption is best for first order of kinetic scheme (so transitions from $C_0$, $C_2$, $O_2$, $D_2$ to $C_1$ - see Scheme 2) and error associated with this approach is $<0.5\%$. For second and fourth order error is on level of $15\%$. The worst accuracy is for the third order of kinetic scheme, for parameters of this model and single exponential decay function of glutamate concentration hitting error level of $40\%$ in a peak of curve. However, in reality the accuracy of all of the orders is better for about factor of 1.8 in peak, due to the fact that glutamate concentration function is not able to perfectly capture dynamics of glutamate concentration in synaptic cleft (see Figure 9). From equation (1), we may see that actually accuracy of the assumption is dependent on glutamate concentration: the bigger glutamate concentration is at PSDs, the better accuracy of the assumption.

\paragraph{New modelling method of temperature effects on AMPA receptor}

Creation of an analytical model for AMPA type synapse, capable to simulate temperature effects, has many potential applications.
Firstly, we achieved model successfully validated with experimental data in closed form analytical solution. The analytical model developed here is capable to include temperature effects with both high accuracy and efficiency in huge neural network simulations, which (when combined with previous studies about voltage-gated ion channels) may open new possibility of researching temperature influence on neural dynamics in computational neuroscience.
Secondly, due to generality\footnote{Sum of multi-exponentials is capable to capture wide-range of various conductance curve dynamics} of this model, it is weakly dependent on kinetic scheme or phenomenological method of synaptic conductance modelling we have chosen (which was discussed as an important problem in the Introduction). This in fact mean that we may generalize some model created in one temperature to any other, by using the model developed here as a 'linking bridge', without performing any additional experiments. This can be done as follows: 1. To some synaptic conductance curve we fit (with free parameters being rate constants and glutamate concentration constants $A$,$w$) the model developed here. 2. Then in fitted model, we multiply all of the AMPAR kinetic rate constants by temperature dependent factors $Q_{10}=2.4$ and hence create new synaptic conductance curve. 3. We fit any phenomenological model to the synaptic conductance curve we achieved in previous step. (see also Scheme 4). 
This approach is insensitive on eventual fitting parameter degeneration, due to the fact of the same multiplication $Q_{10}=2.4$ factor. 

\begin{figure}[h]
\caption{Scheme 4}
\centering
\includegraphics[width=0.5\textwidth]{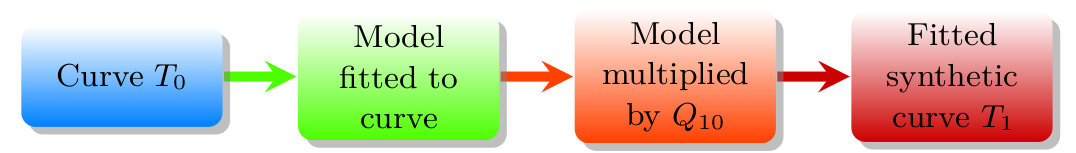}
\end{figure}

The model was efficiently implemented in NEURON \cite{neuron:dg} using NMODL \cite{nmodl:dg}, making it ready for easy implementation in neural network simulations.  

\section{Conclusion}

In the present study, an analytical model of AMPA-type synapse including temperature effects was created. It was done on basis of Markov Models describing kinetics of AMPA receptor, uncoupling of differential equation system method and simplifications motivated by Monte Carlo simulation of synaptic transmission. Due to its generality, model may be used to make any simple model of synaptic conductance easily-scalable for any temperature, which provides simple theoretical linking of research conducted in different temperatures. Thanks to its accuracy (in comparison to experimental data) and efficiency, model may be used in big neural network simulations. This opens new possibility of research various temperature effects on neural dynamics in large-scale multi-neuron experiments and may provide theoretical basis of better understanding of different neurological disorders associated with sub- and super- physiological temperatures. \\
From previous research of shifting attention phenomena \cite{wazny:dg} and large biological neural networks dynamics in function of connections number \cite{wojcik2015:dg}turns out that with large number of varying parameters it may be difficult to find optimal ranges of simulation initial conditions. That is why we may be in demand of using artificial neural network approach (\cite{tadeusiewicz2008:dg}, \cite{tadeusiewicz2013:dg}) in order to find starting points for future investigations of temperature influence on neural dynamics. Such approach may shed some light on understanding paradoxical temperature influence in serious neurological disorders like autism spectrum disorder \cite{autism:dg}, which will be in a scope of our interest in the forthcoming future.

\section{Acknowledgments}

Special thanks to Matthias Hennig, Daniel Wojcik and Piotr Kononowicz, who helped in development of this work.

\begin{landscape}

\section{Appendix A}
Solution of all order of differential equation system describing AMPA receptor kinetics. \\

   $y_{scaled}(t)=0.1 A
   e^{-t (k_c+\omega)} k_o (-\frac{e^{t \omega} S}{R}+\frac{e^{(k_c-P)
   t}}{PR}+\frac{e^{k_c t}}{PS}) k_b +0.4 A^2 e^{-t (k_c+P+2 \omega)} k_o
   (-\frac{e^{t (P+S+\omega)} (k_c-G)}{P R (P-\omega) (S-\omega)}-\frac{e^{t
   (k_c+\omega)} (\omega-S)}{R (P-\omega) (S-\omega) \omega}+\frac{e^{k_c t} (\omega-S)}{P (S-\omega) \omega
   (R+\omega)}+\frac{e^{t (P+2 \omega)} (\omega-P)}{R (P-\omega) (S-\omega) (R+\omega)}) k_b^2+0.35A^3 e^{-t (k_c+P+3 \omega)}k_o \allowbreak
   (-\frac{2 e^{t (P+S+\omega)} (k_c-G) (-G+S-\omega) (S-G)}{P R
   (k_c-3 \omega) (P-2 \omega) (P-\omega) (R+\omega) (R+2 \omega)}-\frac{e^{t (k_c+2 \omega)} (-G+S-\omega)
   (\omega-P) (S-G)}{R (P-2 \omega) (P-\omega) \omega^2 (R+\omega) (R+2 \omega)}+\frac{e^{k_c t}
   (k_c-G) (S-G)}{P R \omega^2 (R+\omega) (R+2 \omega)}-\frac{2 e^{t (P+3 \omega)}}{R (k_c-3
   \omega) (R+\omega) (R+2 \omega)}-\frac{2 e^{t (k_c+\omega)} (k_c-G) (-G+S-\omega)}{R (P-\omega) \omega^2
   (R+\omega) (R+2 \omega)}) k_b^3+A^4 e^{-t (k_c+P+4 \omega)} k_o \allowbreak
(-\frac{e^{t (k_c+3 \omega)} (S-G)(-G+S-\omega) (\omega-P) (3 \omega-G) (-G+k_c-3 \omega)}{6 R (P-3 \omega) (P-2 \omega) (P-\omega) \omega^3 (R+\omega)
   (R+2 \omega) (R+3 \omega)}+\frac{e^{t (P+S+\omega)} (k_c-G) (S-G) (-G+S-\omega) (-G+k_c-3
   \omega)}{P R (k_c-4 \omega) (P-3 \omega) (P-2 \omega) (P-\omega) (R+\omega) (R+2 \omega) (R+3 \omega)}+\frac{e^{t
   (k_c+2 \omega)} (k_c-G) (-G+S-\omega) (-G+k_c-3 \omega)}{2 R (P-2 \omega) \omega^3 (R+\omega)
   (R+2 \omega) (R+3 \omega)}-\frac{e^{t (k_c+\omega)} (k_c-G) (S-G) (-G+k_c-3
   \omega)}{2 R (P-\omega) \omega^3 (R+\omega) (R+2 \omega) (R+3 \omega)}-\frac{e^{t (P+4 \omega)}}{R (k_c-4 \omega)
   (R+\omega) (R+2 \omega) (R+3 \omega)}+\frac{e^{k_c t} (k_c-G) (S-G) (-G+S-\omega)}{6 P R
   \omega^3 (R+\omega) (R+2 \omega) (R+3 \omega)}) k_b^4$
   \newline \newline where, $S=k_c-\omega$, $R=-k_c+k_d+k_o+k_u$, $P=k_d+k_o+k_u-\omega$, $G=k_d+k_o+k_u$
%----------------------------------------------------------------------------------------
%	REFERENCE LIST
%----------------------------------------------------------------------------------------

\end{landscape}

%----------------------------------------------------------------------------------------

\end{document}